\documentclass[reprint,amsmath,amssymb,aps,twocolumn,superscriptaddress]{revtex4}

\usepackage{graphicx}
\usepackage{dcolumn}
\usepackage{mathrsfs}
\usepackage{xcolor}
\usepackage{ulem}
\usepackage{appendix}
\usepackage{enumitem}
\newlist{todolist}{itemize}{2}
\setlist[todolist]{label=$\square$}

\begin{document}
\title{Thermodynamics of microphase separation in a swollen, strain-stiffening polymer network}

\author{Carla Fern\'{a}ndez-Rico$^*$}
\affiliation{Department of Materials, ETH Z\"{u}rich, 8093 Zurich, Switzerland}

\author{Robert W. Style$^*$}
\affiliation{Department of Materials, ETH Z\"{u}rich, 8093 Zurich, Switzerland}
\author{Stefanie Heyden}
\affiliation{Department of Materials, ETH Z\"{u}rich, 8093 Zurich, Switzerland}

\author{Shichen Wang}
\affiliation{Department of Physics, Georgetown University, Washington DC, USA}

\author{Peter D. Olmsted}
\affiliation{Department of Physics, Georgetown University, Washington DC, USA}

\author{Eric R. Dufresne}
\affiliation{Department of Materials Science and Engineering, Department of Physics, Cornell University, Ithaca, NY 14853, USA}
\affiliation{Department of Materials, ETH Z\"{u}rich, 8093 Zurich, Switzerland}

\date{\today} 

\begin{abstract}
Elastic MicroPhase Separation (EMPS) provides a simple route to create soft materials with homogeneous microstructures by leveraging the supersaturation of crosslinked polymer networks with liquids.
At low supersaturation, network elasticity stabilizes a uniform mixture, but beyond a critical threshold, metastable microphase-separated domains emerge.
While previous theories have focused on describing  qualitative features about the size and morphology of these domains, they do not make quantitative predictions about EMPS phase diagrams.
In this work, we extend Flory-Huggins theory to quantitatively capture EMPS phase diagrams by incorporating strain-stiffening effects. 
This model requires no fitting parameters and relies solely on independently measured solubility parameters and large-deformation mechanical responses. Our results reveal that strain-stiffening enables metastable microphase separation within the swelling equilibrium state and why the microstructures can range from discrete droplets to bicontinuous networks. This works highlights the critical role of nonlinear elasticity in controlling phase-separated morphologies in polymer gels.
\end{abstract}


\maketitle

\section{Introduction}

Phase separation is a ubiquitous process that can be harnessed to create structured materials \cite{xu2023liquidRev,wang2024constructingRev}.
In this process, two or more components demix, typically triggered by changes in temperature, composition, or degree of polymerization \cite{CompInducedPS2011,TempInducedPS19911,fily2012athermal}.
Demixing creates microstructures with discrete domains or bicontinuous channels.
For classical phase-separation processes, like liquid-liquid phase separatation, domains  gradually coarsen due to the influence of surface tension \cite{gibbs1879equilibrium,Jones2002}.
When this coarsening is arrested (for example by solidification, or going through a glass transition \cite{manley2005glasslike,wienk1996recent,cardinaux2007interplay}), solid materials with intricate and stable microstructures can be created \cite{FernandezRico2022,sicher2021structural}.
Synthetic examples include metal alloys \cite{ardell1985precipitation}, membranes \cite{Guillen_2011_Nonsolvent}, and cell tissue scaffolds \cite{dudaryeva2024tunable}, while natural examples include structurally-colored materials produced by many different animals \cite{burg2018self,dufresne2009self}, and phase-separated domains inside cytoplasm and lipid membranes \cite{brangwynne2009germline,Feric2016,banani2017biomolecular,elson2010phase}.

\begin{figure}[t]
\includegraphics[width=\columnwidth]{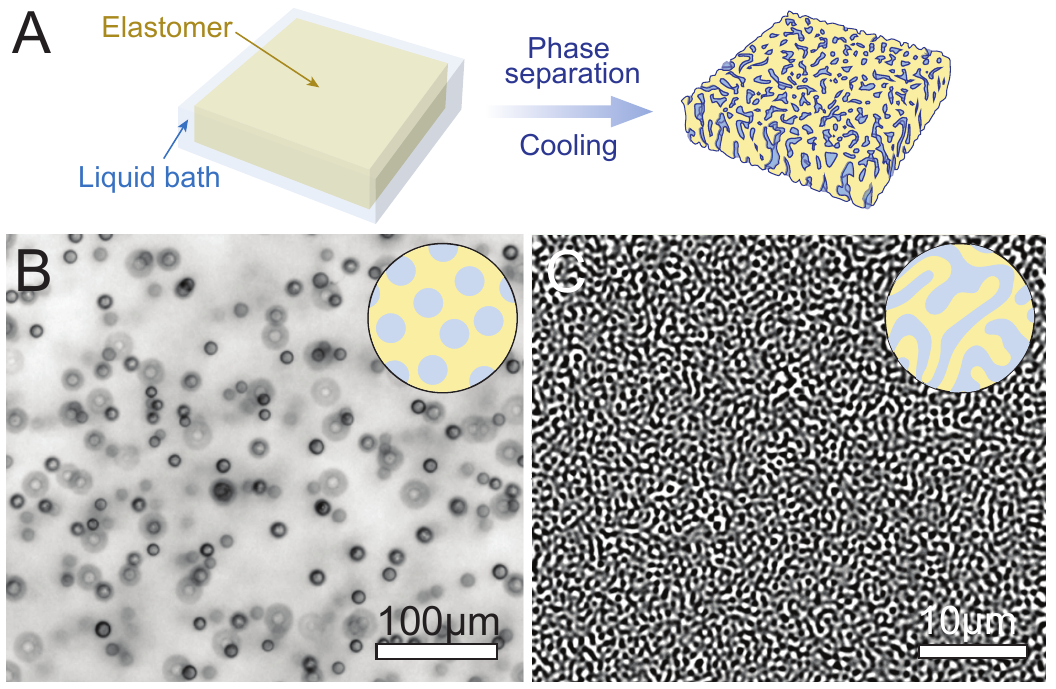}
	\caption{Schematic of EMPS. A) First, an elastomer is swollen in a liquid bath at elevated temperatures. After swelling equilibrium is reached, cooling induces phase separation. Different phase separated morphologies and sizes are obtained as a function of the matrix's composition and stiffness: B) droplet \cite{style2018liquid} and C) bi-continuous morphologies \cite{fernandez2024elastic}}
	\label{fig:Figure1}
\end{figure}

Recently, Elastic Microphase Separation (EMPS) has emerged as an alternative, convenient route to create phase-separated microstructures that do not coarsen \cite{fernandez2024elastic,style2018liquid,fernandez2021putting}. In essence, EMPS exploits mechanical forces imposed by a polymer network to counteract interfacial tension.
In typical EMPS experiments, an elastomer is equilibrated with solvent at high temperatures (see Figure 1A). Then, cooling triggers phase separation of the solvent within the elastomer. 
Phase separation occurs rapidly and locally, within a few seconds.
Microstructures with either discrete droplets or bicontinuous channels form, depending on the elastomer's  degree of swelling and stiffness (see Figure \ref{fig:Figure1}BC).
This process is simple and robust enough to easily produce large pieces of material with precisely controlled microstructures. 
While these microphase-separated structures are stable to coarsening, they are unstable to the loss of solvent to the surrounding bath on a timescale that is $O(10\, \mathrm{hrs})$ \cite{rosowski2020elastic}.
The onset of elastic microphase separation is characterized by a novel phase boundary inside the super-saturated regime of the network-solvent swelling equilibrium, as described in references \cite{fernandez2024elastic, style2018liquid}.

EMPS is closely connected to liquid-liquid phase separation.
At the monomer scale, both the elastomer and solvent are liquid-like.
Thus, we expect the earliest stages of phase separation to be insensitive to the relatively sparse elastic constraints imposed by the cross-links between the polymer chains.
However, at later stages of phase separation, we expect the elasticity of the polymer phase to become important.
Indeed experiments suggest that the elastomer's Young's modulus plays an essential role in defining morphology and thermodynamics \cite{fernandez2024elastic,style2018liquid,rosowski2020elastic,shin2018liquid}.
Importantly though, theoretical studies agree that linear elasticity is insufficient to capture key aspects of the experiments. 
For example, previous works have investigated the role of mesh size \cite{ronceray2022liquid,liu2023liquid}, viscoelasticity \cite{tanaka2000viscoelastic}, damage \cite{vidal2021cavitation,paulin2022fluid},  and heterogeneous or non-local elasticity \cite{qiang2024nonlocal,vidal2020theory,meng2024heterogeneous,mannattil2024theory,deviri2024mechanosensitivity} and   non-linear elasticity \cite{hennessy2020phase,biswas2022thermodynamics,ronceray2022liquid}.

Out of all these topics, several theoretical works have highlighted that strain stiffening of the polymer network should be a key feature controlling EMPS.
For example, strain-stiffening has been predicted to stabilize microphase-separated structures against ripening \cite{biswas2022thermodynamics}, while also controlling the size of the microdomains \cite{kothari2020effect,wei2020modeling}.
Furthermore,  \citet{ronceray2022liquid} have shown that strain-stiffening plays a key role in determining whether phase microphase-separated domains stay smaller than the elastomer's mesh size, or whether they are able to grow to more macroscopic sizes.
Finally,  \citet{hennessy2020phase} showed that 1D gels with non-linear elasticity have similar phase diagrams to EMPS with two distinct phase boundaries.
While all of the above theories qualitatively capture  different aspects of EMPS, there have been no quantitative comparisons of theory and experiments.

Here, we explore the impact of strain-stiffening on the phase separation of liquids within elastic matrices by  extending Flory-Huggins' (FH) theory of macromolecular liquid-liquid phase separation, incorporating energy stored in large elastic deformations. 
We show that strain-stiffening enables  metastable microphase separation in a subset of the super-saturated region of the swelling phase diagram.
Using this model, we can quantitatively predict EMPS phase diagrams from measurements of the i)  stress-strain relationship of the  neat elastomer and ii) solubility of uncrosslinked polymer in the solvent. 
These predictions are in reasonable agreement with measured phase diagrams of fluorinated oil in a silicone elastomer. 
Finally, we discuss the role of cavitation instabilities in selecting the microphase-separated morphology.

\section{Liquid-liquid phase separation}

\begin{figure}
\centering
\includegraphics[width=0.65\columnwidth]{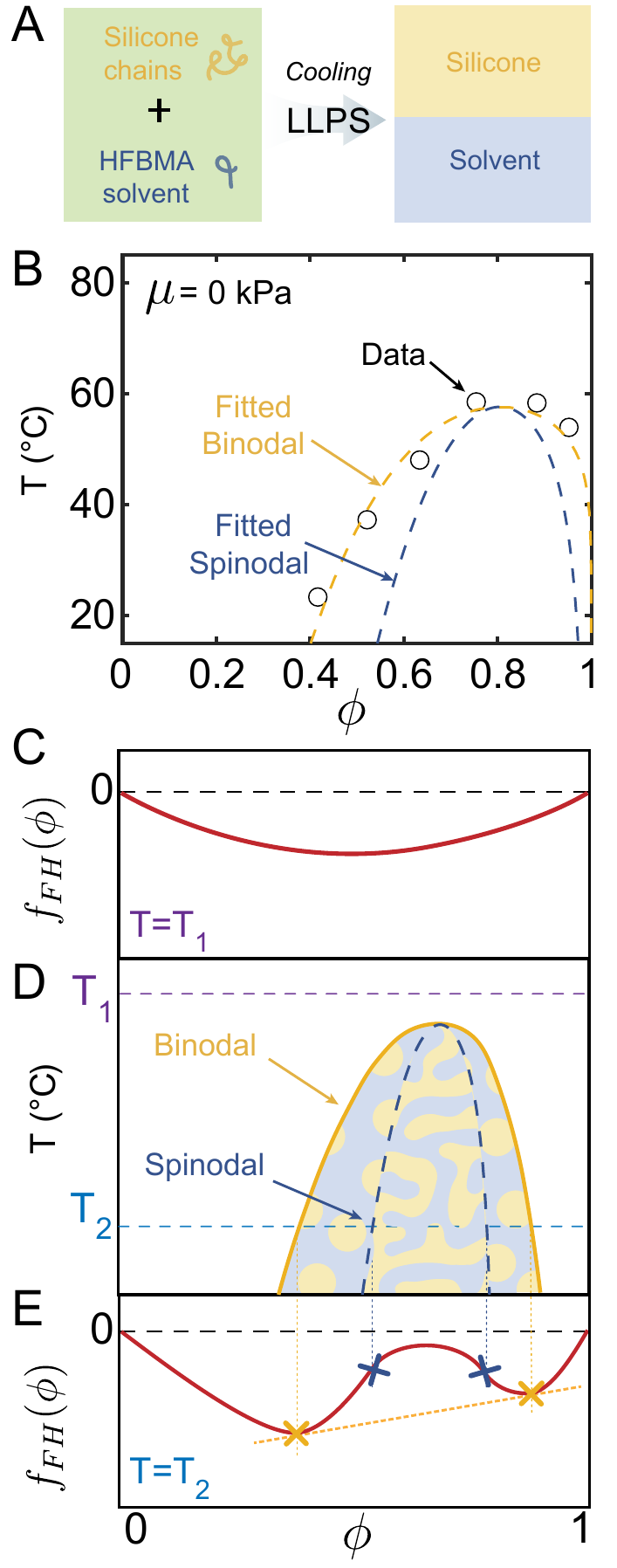}
	\caption{Liquid-liquid phase separation behavior of uncrosslinked silicone polymer mixed with HFBMA solvent. A) Schematic of macroscopic phase separation.
    B) Experimental phase diagram of the LLPS silicone - HFBMA (see black circles) fitted with classical FH theory (see dashed yellow line). Here, $\phi$ is the solvent volume fraction. C,E) Schematic of the total free energy of mixing at different temperatures, $T_1$ and $T_2$. D) Schematic LLPS diagram corresponding to C,E.}
	\label{fig:LLPS}
\end{figure}

In order to understand how elasticity affects phase separation, we start by characterizing  liquid-liquid phase separation of the elastomer's uncrosslinked polymer precursors in the solvent of interest. 
Here, this consists of 28 kDa vinyl-terminated PDMS chains in heptafluorobutyl methacrylate (HFBMA, 268.1 Da) (see Figure \ref{fig:LLPS}A).
As reported previously \cite{fernandez2024elastic}, we determine the phase diagram of this system by recording the temperature at which phase separation initiates during simultaneous cooling and observation under a light microscope (see Materials \& Methods). The resulting phase diagram takes the classical LLPS form shown in Figure \ref{fig:LLPS}B (see black circles).
Above a critical temperature $T_c \approx 60^\circ$C, the two fluids are always miscible.
At lower temperatures, demixing occurs for a range of HFBMA volume fractions, $\phi$, in the mixture.

LLPS phenomena can be described using Flory-Huggins (FH) theory \cite{flory1942thermodynamics}, which expresses the stability of solvent-polymer mixtures based on an expression for the dimensionless Helmholtz free energy per unit volume of a mixture of the form \cite{doi2013soft}:
\begin{equation}
f_\mathrm{FH} (\phi,T) = \phi \log\phi +\frac{v_s}{v_p} (1-\phi) \log(1-\phi) + \chi(\phi,T) \phi (1-\phi),
\label{eq:FH}
\end{equation}
where $\phi$ is the volume fraction of the solvent,  $v_{s}$ and $ v_p$ are the molecular volumes of solvent and polymer, respectively, and $\chi(\phi,T)$  is the Flory-Huggins polymer-solvent interaction parameter \cite{koningsveld2001polymer,pgdgpolymer}. 
The free energy density  $f_{\textrm{FH}}$ is normalized by  $k_{\textrm{B}}T/v_s$, where $k_{\textrm{B}}$ is Boltzmann's constant and $T$ is the  absolute temperature.
The first two terms in this equation represent the entropy of mixing of the two components, which favours a single mixed phase.
The last term represents the enthalpy of mixing, which favours separation into polymer-rich and solvent-rich phases. 

To convert an expression for free energy, like that above, into a phase diagram, we use the classical approach illustrated in Figures  \ref{fig:LLPS}C-D \cite{doi2013soft}.
Figures \ref{fig:LLPS}C and E show schematic representations of $f_{\textrm{FH}}$ at different temperatures (Figure \ref{fig:LLPS}C and E). 
Above $T_c$, $f_\mathrm{FH}(\phi)$ is globally convex downward, due to the dominance of the entropic terms (see Figure  \ref{fig:LLPS}C). The free energy of the mixture can not be lowered by separating into two phases.  Thus, the system remains completely mixed.
Below $T_c$, the enthalpy of mixing becomes more important, and $f_\mathrm{FH}$ develops concavity (see Figure \ref{fig:LLPS}E). 
In this case, phase separation can occur because the system's free energy can be reduced by splitting into two different phases.
The range of $\phi$ where phase separation occurs can be found by applying the double-tangent construction (see full derivation in Appendix A), which essentially consists of a tangent line (see yellow line in Figure  \ref{fig:LLPS}E) that connects the bottom sides of the two convex regions. This double-tangent line ensures that the chemical potentials of the solvent and the polymer are equal in both phases, fulfilling the condition for phase equilibrium. Furthermore, the points of tangency (yellow crosses in Figure \ref{fig:LLPS}E) then represent the equilibrium composition of the two phases and are known as the \textit{binodal} points.

The range of $\phi$ over which the system undergoes phase separation can be further subdivided, based on the curvature of $f_\mathrm{FH}$.
When $\partial^2f_\mathrm{FH}/\partial \phi^2>0$ (outside the blue crosses), the mixture is metastable and phase separation occurs by localized nucleation and growth (see droplet morphologies in Figure \ref{fig:LLPS}D).
When $\partial^2f_\mathrm{FH}/\partial \phi^2<0$,  (between the blue crosses, Figure \ref{fig:LLPS}E), the mixture is unstable, and system separates continuously through  spinodal decomposition (see interconnected structures in Figure \ref{fig:LLPS}D) \cite{doi2013soft}.
By performing this calculation across a range of different temperatures, we construct the binodal and spinodal curves of the phase diagram shown in Figure \ref{fig:LLPS}D).

To fit our LLPS data using Flory-Huggins theory, we need to determine $v_s/v_p$ and $\chi(\phi,T)$. 
While $v_s/v_p$ is easily calculated from molar volumes of the two components ($v_s/v_p=6.90 \times 10^{-3}$),
we fit the experimental binodal (black circles in Figure \ref{fig:LLPS}A) with $\chi(\phi,T)$ \cite{wolf2011making} using a commonly assumed form:
\begin{equation}
    \chi(\phi,T)=A+B/T + C\phi.
    \label{eq:g(T)}
\end{equation}
We find best-fit values of $A=0.2108$, $B=247.76\,\mathrm{K}$ and $C=-0.227$ (see fitted binodal and spinodal dashed lines in Figure \ref{fig:LLPS}B). 

Although it may seem excessive to use three fitting parameters to fit 6 data points, it is not possible to fit the data with less, as each parameter has a separate effect on the phase diagram: 
$A$ essentially moves the phase diagram up and down, $B$ changes the width of the peak, while $C$ is the only parameter that moves the peak laterally. 

In Section \ref{sec:phase_separation_in_a_strain_stiffening_gel}, we will use these parameters to predict EMPS phase diagrams. This assumes that
 $\chi(\phi,T)$ does not change when the silicone chains cross-link into an elastic network. 

\section{Phase Separation in an Elastic Network}

In this Section, we briefly review the essential features of phase diagrams in elastic microphase separating systems before extending Flory-Huggins theory to account for large-deformation elasticity of polymer matrices upon swelling and phase separation.

\begin{figure}[t]
	\includegraphics[width=\columnwidth]{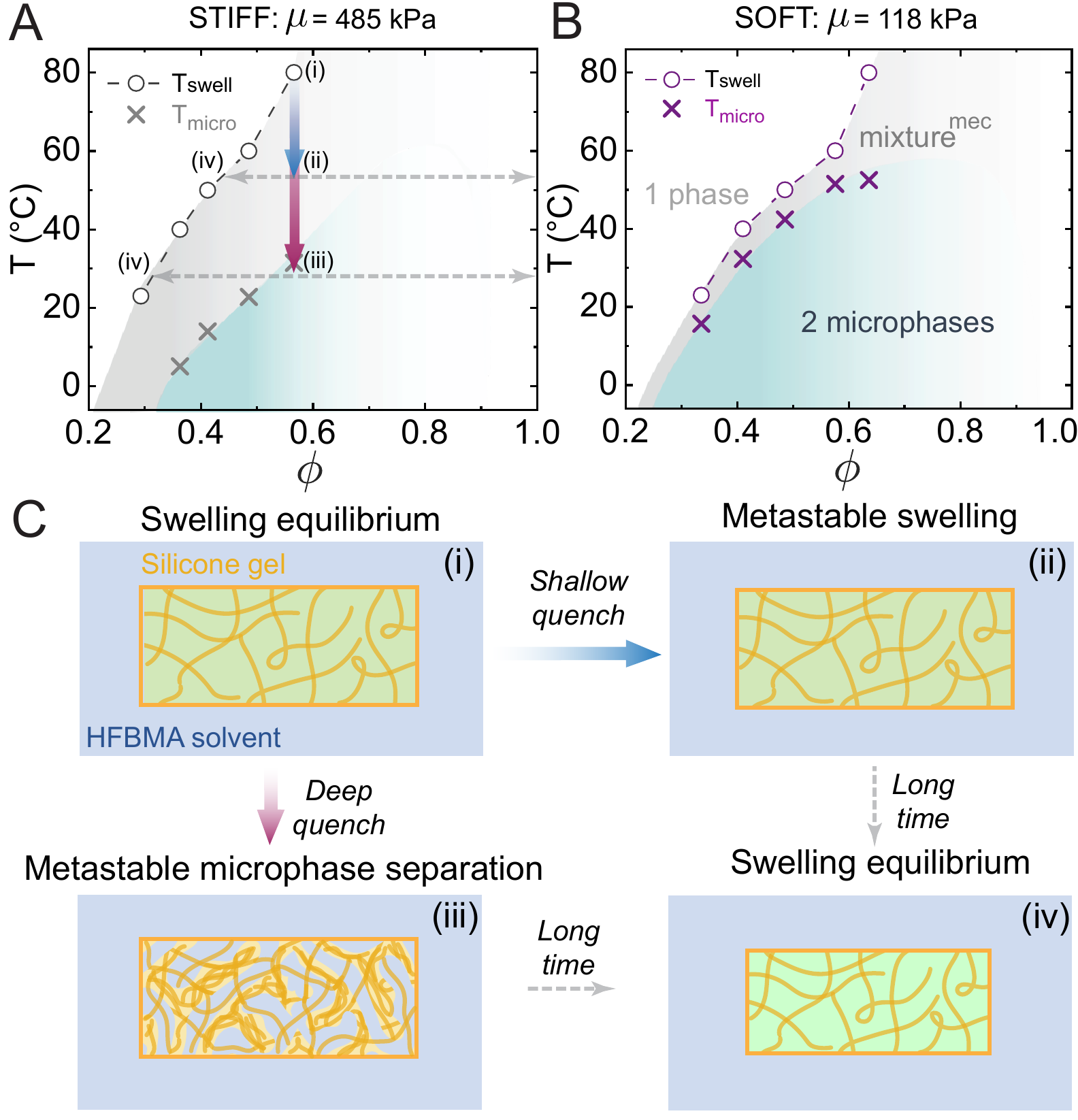}
    \centering
	\caption{ EMPS of HFBMA solvent in silicone polymer matrices. Experimental phase diagrams of the HFBMA - PDMS system for (A) 485kPa (stiff) and (B) 118kPa (soft) samples. The boundaries at $T_{swell}$, $T_{micro}$ are shown by empty circles and x's respectively. The gray area shows the part of the phase diagram where the mixture is stabilized mechanically (mixture $^{mec}$) and the turquoise area where microphase separation is observed (2 microphases). The mixture $^{mec}$ region size increases with $\mu$. C) Schematics of the different states of the system depending on the temperature: (i) Swelling equilibrium, (ii) Metastable swelling, (iii) Metastable microsphase separation, (iv) new swelling equilibrium. 
    }
	\label{fig:EMPS_phase_diagrams}
\end{figure}

Compared to LLPS, crosslinking silicone chains into elastic networks qualitatively changes phase separation of HFBMA in silicone.
This is shown in the experimental phase diagrams for polymer networks with shear moduli $\mu=$ 485 kPa and 118 kPa shown in Figures  \ref{fig:EMPS_phase_diagrams}A,B  respectively. 
The higher boundary (circles) represents macroscopic swelling equilibrium of the silicone swollen in pure HFMBA (see sketch Figure  \ref{fig:EMPS_phase_diagrams}C(i)). 
Below this boundary (shallow quench), the system is super-saturated, but  no microscopic phase separation is observable (see Figure \ref{fig:EMPS_phase_diagrams}C(ii)).
Instead,  the solvent inside the matrix diffuses to the bath until it reaches a new swelling equilibrium (see Figure \ref{fig:EMPS_phase_diagrams}C(iv)). This deswelling process takes a time $\tau_{deswell}= O(\mathrm{10\,hrs})$ for a cm-scale sample.
The lower boundary (crosses) marks the onset of microphase separation. 
When cooled below this line (deep quench), the system undergoes microscopic phase separation as shown in Figure  \ref{fig:EMPS_phase_diagrams}C(iii).
This microscopic process occurs within a much shorter time, $\tau_{\mu PS}=O(1\mathrm{\,min})$.
The micro-scale structures are metastable and after a long time, $\sim \tau_{deswell}$, the solvent inside the microstructures diffuses to the bath and reaches a new swelling equilibrium (see dashed arrows).
The microphase-separation boundary does not behave like  classical spinodal and binodal lines curve. Like a spinodal, we observe a continuous phase transformation when the boundary is crossed. Like a binodal, phase-separated domains dissolve at the same temperature they form \cite{fernandez2024elastic}.

Importantly, the size of the gap between the swelling equilibrium and microphase boundaries depends on the stiffness of the unswollen silicone, with the stiffer silicone having a much larger gap than the softer one \cite{fernandez2024elastic}. The gap, or `mechanically-stabilized regime', vanishes as stiffness decreases, and the two phase boundaries merge into one that approach the LLPS phase boundary in Figure~\ref{fig:LLPS}B.

There are two further key differences between phase separation in an elastic network and LLPS. 
First, while LLPS has a critical temperature, swollen elastic networks will always phase separate  at sufficiently high solvent concentrations.  
Second, while phase-separated domains in LLPS always coarsen (due to coalescence and Ostwald ripening) \cite{meng2024heterogeneous,tanaka2000viscoelastic}, this coarsening is completely suppressed in EMPS \cite{style2018liquid,fernandez2024elastic,rosowski2020elastic}.

\subsection{Thermodynamics of phase separation in an elastic network \label{sec:thermo_emps}}

To model these phenomena, we update the free-energy density shown in Eq.~(\ref{eq:FH}) to include the effects of crosslinking of the silicone chains into an elastic network.
Crosslinking removes the translational entropy of the polymer (i.e. the second term in Eq.~1) as $v_p/v_s\rightarrow \infty$.
We account for elasticity of the polymer network by simply adding an additional elastic free-energy density $f_{el}$ to $f_\mathrm{FH}$ \cite{mckenna1990swelling,hong2008theory,flory1943statistical,wei2020modeling,hennessy2020phase}.
Then, the resulting free-energy density is
\begin{equation}
    f_\mathrm{tot}(\phi,T)= \phi \log\phi + \chi(\phi,T) \phi (1-\phi)+f_\mathrm{el}(\phi)
    \label{eq:f_tot}.
\end{equation}
For comparison with data, we will assume that crosslinking does not alter the form of $\chi(\phi,T)$  (Eq.~\ref{eq:g(T)}).
Here, $f_\mathrm{el}$ is related to the standard strain-energy density, $W_\mathrm{el}$, from elasticity theory, as:
\begin{equation}
f_\mathrm{el}=(1-\phi) \frac{W_\mathrm{el}}{k_{\textrm{B}}T/v_s}.
\label{eq:FEL} 
\end{equation}
Here, the factor of $k_{\textrm{B}}T/v_s$ non-dimensionalizes $W_\mathrm{el}$, while the factor of $(1-\phi)$ converts from a free energy per volume of dry polymer to a free energy per unit volume in the swollen state.
In general, $W_\mathrm{el}$ is a function of the principal stretches in the polymer network: $\lambda_1$, $\lambda_2$, $\lambda_3$.
For many simple polymer networks \cite{binder1984phase}, 
\begin{equation}  W_\mathrm{el}=\tilde{W}_\mathrm{el}(I_1)-\alpha \mu \log J,
\label{eq:alpha}
\end{equation}
 where $I_1=\lambda_1^2+\lambda_2^2+\lambda_3^2$ is the mean-squared stretch in the polymer network, $J=\lambda_1\lambda_2\lambda_3$ is the change in volume of the network relative to the dry state, $\mu$ is the shear modulus, and $\alpha$ is a positive dimensionless number. 
The second term on the right-hand side reflects the increased entropy of the crosslinking points during swelling.
Here, we will assume isotropic swelling, $\lambda_i=(1-\phi)^{-1/3}$, so that $I_1=3(1-\phi)^{-2/3}$ and $J=(1-\phi)^{-1}$, and initially take $\alpha=0$ \cite{james1953statistical}.

\subsection{Characterizing polymer network elasticity}

\begin{figure}[h]
	\centering
    \includegraphics[width=0.85\columnwidth]{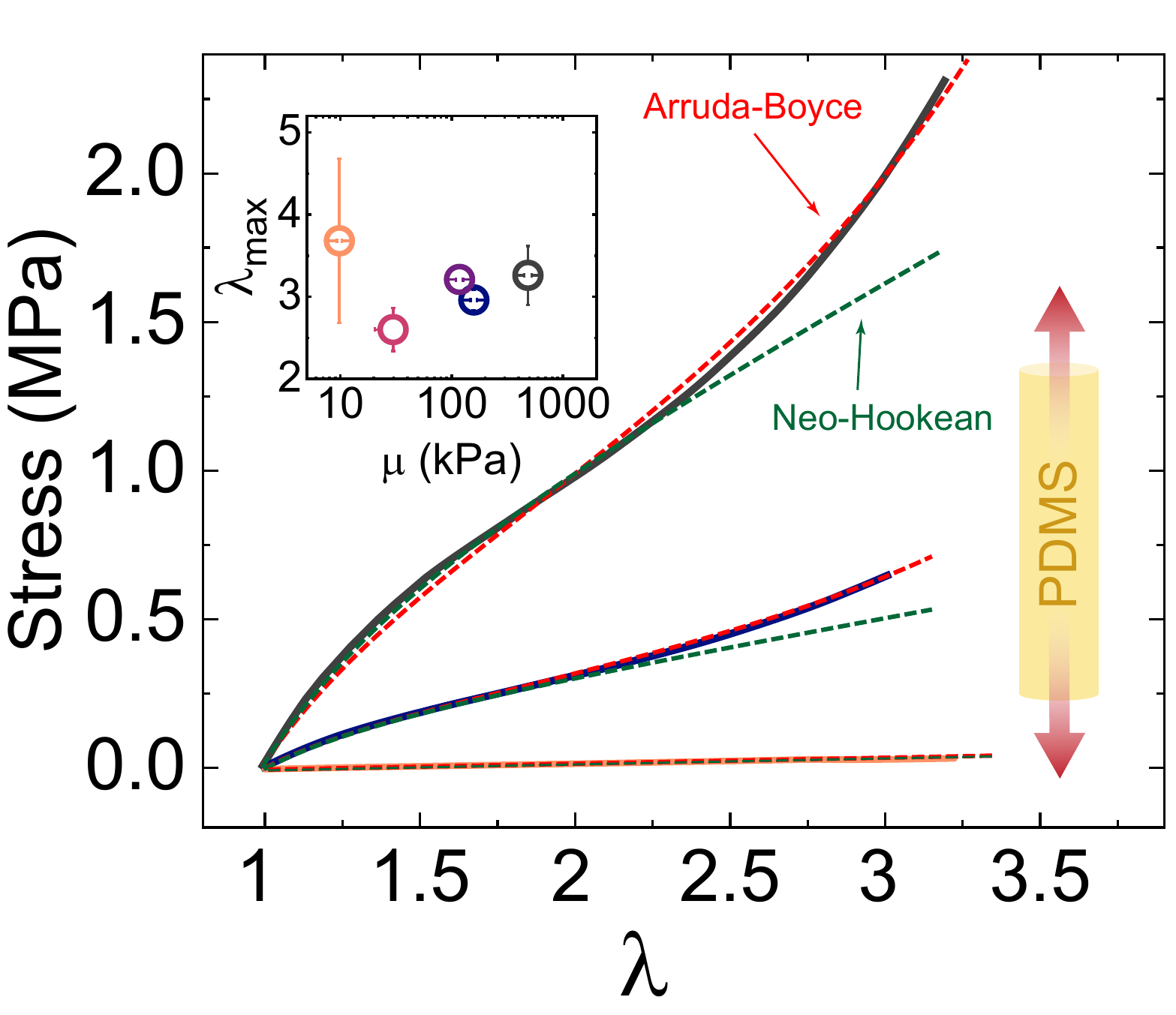}
	\caption{Mechanical characterization of the silicone polymer networks prior to swelling with the HFBMA solvent. A) Examples of  engineering stress versus stretch $\lambda$ for an incompressible uniaxial extension, with neo-Hookean curves fitted up to $\lambda=2$ and then Arruda-Boyce curves fit over the whole range of $\lambda$. The inset shows the Arruda-Boyce fit parameters for all the materials used here (see Table~\ref{table} in Materials and Methods). Error bars show standard deviations of the measured values.
    }
	\label{fig:elasticity}
\end{figure}

We obtain expressions for $\tilde{W}_\mathrm{el}(I_1)$ for our polymer networks by performing large extension tensile tests on unswollen samples ($\phi=0,\,\,J=1$) of the five different types of silicone elastomers used in our EMPS experiments (see Materials and Methods).
Example plots of engineering stress, $\sigma^\mathrm{eng}$, versus stretch, $\lambda$, are shown in Figure \ref{fig:elasticity} for several different stiffness silicone networks.

At small stretch ($\lambda\lesssim 2$), all the polymer networks can be well-described by classical (non strain-stiffening) neo-Hookean solids (see fits as green dashed lines in Figure \ref{fig:elasticity}), with
\begin{equation}
\tilde{W}_\mathrm{el}^\mathrm{nh}(I_1)=\frac{\mu}{2}(I_1-3),
\label{eq:W_NH}
\end{equation}
where $\mu$ is the small-strain shear modulus.
For uniaxial tension, the relation between engineering stress and stretch  is calculated from $\tilde{W}_\mathrm{el}$ as \cite{marckmann2006comparison}:
\begin{equation}
    \sigma^\mathrm{eng} = 2\left(\lambda-\frac{1}{\lambda^2} \right)\left.\frac{d \tilde{W}_\mathrm{el}}{d I_1}\right|_{I_1=\lambda^2+2/\lambda},
    \label{eq:sed}
\end{equation}
where $\lambda$ is the uniaxial stretch, and we assume that the dry polymer network is incompressible.

For $\lambda>2$, all our silicone elastomers strain stiffen, diverging from the neo-Hookean model predictions (see fits as green dashed lines in Figure \ref{fig:elasticity}).
To capture this behaviour, we fit the results with the strain-stiffening Arruda-Boyce model \cite{arruda1993three}.
This continuum model is based on the physics of the underlying polymer chains, 
accounting for the limiting stretch of each chain.
The Arruda-Boyce model can be approximated by
\begin{equation}
    \tilde{W}_\mathrm{el}^\mathrm{ab}=\frac{\mu}{2}\frac{\sum_{i=1}^5 \frac{\alpha_i}{\lambda_m^{2i-2}}(I_1^i-3^i)}{\sum_{i=1}^5 \frac{i\alpha_i}{\lambda_m^{2i-2}}3^{i-1}},
    \label{eq:W_AB}
\end{equation}
where $\alpha_1=1/2$, $\alpha_2=1/20$, $\alpha_3=11/1050$, $\alpha_4=19/7000$ and $\alpha_5=519/673750$.
Again, $\mu$ is the small-strain shear modulus, while $\lambda_m$ is the maximum ``locking" stretch of individual polymer chains relative to their end-to-end distance in the as-prepared elastomers. 
The stress-stretch relationship is again found by inserting the strain-energy density into Eq.~(\ref{eq:sed}).
The resulting fits are shown in Figure \ref{fig:elasticity} as the red dashed lines, showing a good agreement between the data and the model over the whole range of $\lambda$. We extract values of $\mu$ and $\lambda_m$ for five different polymer networks containing different amounts of crosslinker, and obtain values of $\mu$ ranging from 10 to $\sim$ 490kPa and $\lambda_m$ values ranging from 4 and 3 (see Table 1, Materials and Methods), as shown in the inset in Figure \ref{fig:elasticity}.
Note that in Figure 4, the samples reach  stretches   $\lambda>\lambda_m$.
This is counterintuitive, but permitted in such uniaxial tests, as individual chains stretch significantly less than the applied macroscopic stretch $\lambda$ applied to a sample.
\textit{E.g.}, in the Arruda-Boyce model, individual chain stretch during a uniaxial test is $\sqrt{(\lambda^2+2/\lambda)/3}$ \cite{rickaby2015comparison}.
Thus, chain locking (which causes stress divergence) will only occur when this reaches $\lambda_m$.
\textit{E.g.} for a material with $\lambda_m=3$, the maximum possible uniaxial sample stretch is $\lambda=5.2$.

Note that in previous works \cite{fernandez2024elastic,style2018liquid} we characterized the same silicone samples using indentation with a flat punch.
Here, we characterize the materials using uniaxial tension testing, as this allows us to probe the large-strain material response, and provides more accurate measurements.
We also use $\mu$ instead of the Young's modulus $E$ (which we reported previously), as this is a more natural parameter for working with swollen materials \cite{rice1976some}.

With the mechanical fits, we now have expressions for $f_\mathrm{el}$ that we can use in the total free energy (\ref{eq:f_tot}).
This allows us to compare the predicted phase diagrams for the neo-Hookean and Arruda-Boyce models, and  examine the role of strain-stiffening in EMPS.

\subsection{Phase separation in a neo-Hookean matrix}

\begin{figure}
	\centering
    \includegraphics[width=0.6\columnwidth]{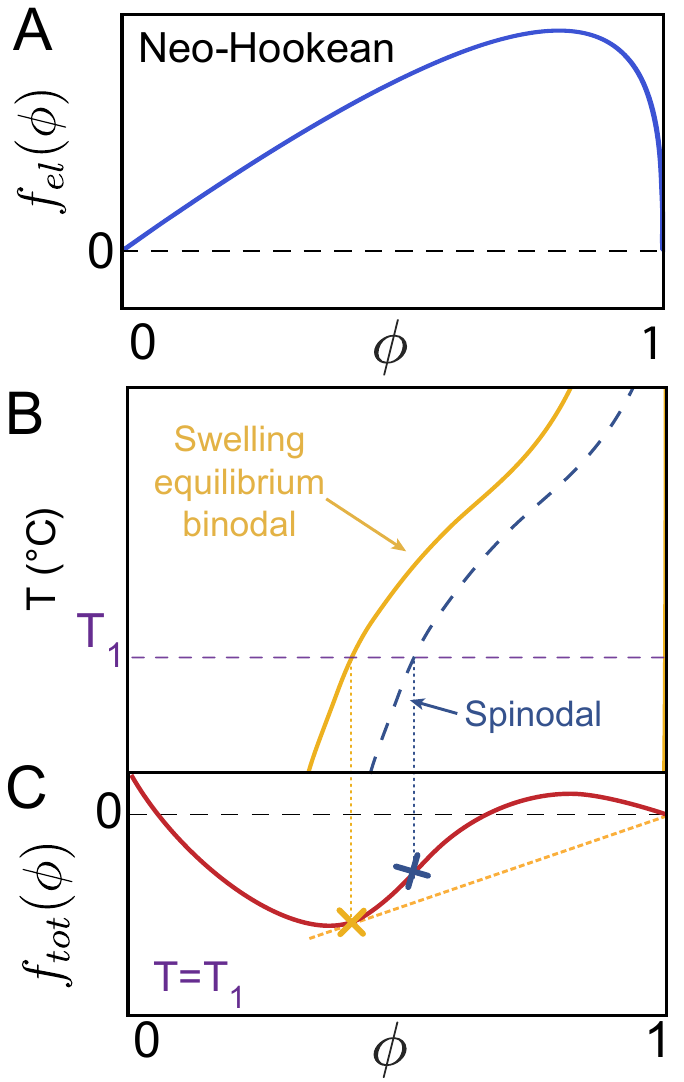}
	\caption{Effect of neo-Hookean elastic energy on the classical FH free energy of mixing. A) Schematic of the elastic contribution to the free-energy density for the neo-Hookean model as a function of solvent volume fraction, $\phi$. B) Schematic phase diagram. C) Schematic of the total energy of mixing for a neo-Hookean system.}
	\label{fig:nh_phase_diagram}
\end{figure}

We first examine the stability of swollen neo-Hookean networks. This is the simplest hyperelastic model, and does not feature any strain stiffening. 
To do this, we insert Eq.~(\ref{eq:W_NH}) into Eq.~(\ref{eq:FEL}) to obtain
\begin{equation}
f_\mathrm{el}^\mathrm{nh}=3\bar{\mu}\left[\left(1-\phi\right)^{1/3}-(1-\phi)\right],
\label{eq:f_elnh}
\end{equation}
where $\bar{\mu}=\mu /(2k_{\textrm{B}}T/v_s)$ is the dimensionless, small-strain shear modulus.
The form of $f_\mathrm{el}^\mathrm{nh}(\phi)$ is shown in Figure \ref{fig:nh_phase_diagram}A.
 Interestingly, it does not monotonically increase as the network. 
This is because the network volume increases faster than the total stored elastic energy in the system.
Thus, the elastic energy density (measured with respect to the current state) decreases at high swelling.

To calculate the phase diagram for EMPS, where we ignore the slow exchange of solvent with the surroundings,  we must include   $f_\mathrm{el}^\mathrm{nh}(\phi)$ in the free energy, Eq.~(\ref{eq:f_tot}) (see full expression in Appendix B).
Then, $f_\mathrm{tot}(\phi)$ takes the form shown schematically in Figure \ref{fig:nh_phase_diagram}C, where the curve is convex for small $\phi$, and concave for large $\phi$ (where elasticity dominates).
This concave section indicates the presence of phase separation, but  there is no stabilizing convexity at large $\phi$, which is needed to perform the double-tangent construction. Hence, thermodynamic phase separation between networks at different levels of swelling (\textit{i.e.} EMPS) is not possible using neo-Hookean elasticity and the measured $\chi(\phi,T)$. 

However, there is a  macroscopic swelling equilibrium between pure HFMBA and a swollen network, which occurs at long times.
Geometrically, this is found with a single-tangent construction that  passes through the point where $(f_\mathrm{tot}=0,\phi=1)$ (see dotted line in the Figure, and derivation in  Appendix A).
The tangent point (yellow cross) then represents swelling equilibrium in pure solvent, while spinodal decomposition still happens when $\partial^2 f_\mathrm{tot}/\partial \phi^2<0$ (to the right of  the blue cross).
The resulting schematic form of the phase diagram is shown in Figure \ref{fig:nh_phase_diagram}B.
Note that the swelling equilibrium curve is essentially the same as the predictions of Flory-Rehner theory \cite{flory1943statistical}. 

Overall, the resulting phase diagram does not fully capture the EMPS behaviour seen in Figure \ref{fig:EMPS_phase_diagrams}.
While this model is capable of describing the swelling equilibrium boundary (long-time equilibrium at fixed chemical potential, with solvent exchange between the gel and the surroundings) it cannot predict EMPS (short-time phase separation at fixed volume).
Note that the spinodal boundary (see dashed boundary in Figure \ref{fig:nh_phase_diagram}B) cannot explain the onset of EMPS.
If phase separation were to occur at the spinodal, the resulting phase-separated domains would only dissolve when heated up to the separate binodal curve. However, when observing EMPS in experiments, we see the formation and dissolution of microphase separated domains at the same temperature \cite{fernandez2024elastic}.

\subsection{Phase separation in a strain-stiffening matrix}
\label{sec:phase_separation_in_a_strain_stiffening_gel}

\begin{figure}[t]
	\centering\includegraphics[width=0.6\columnwidth]{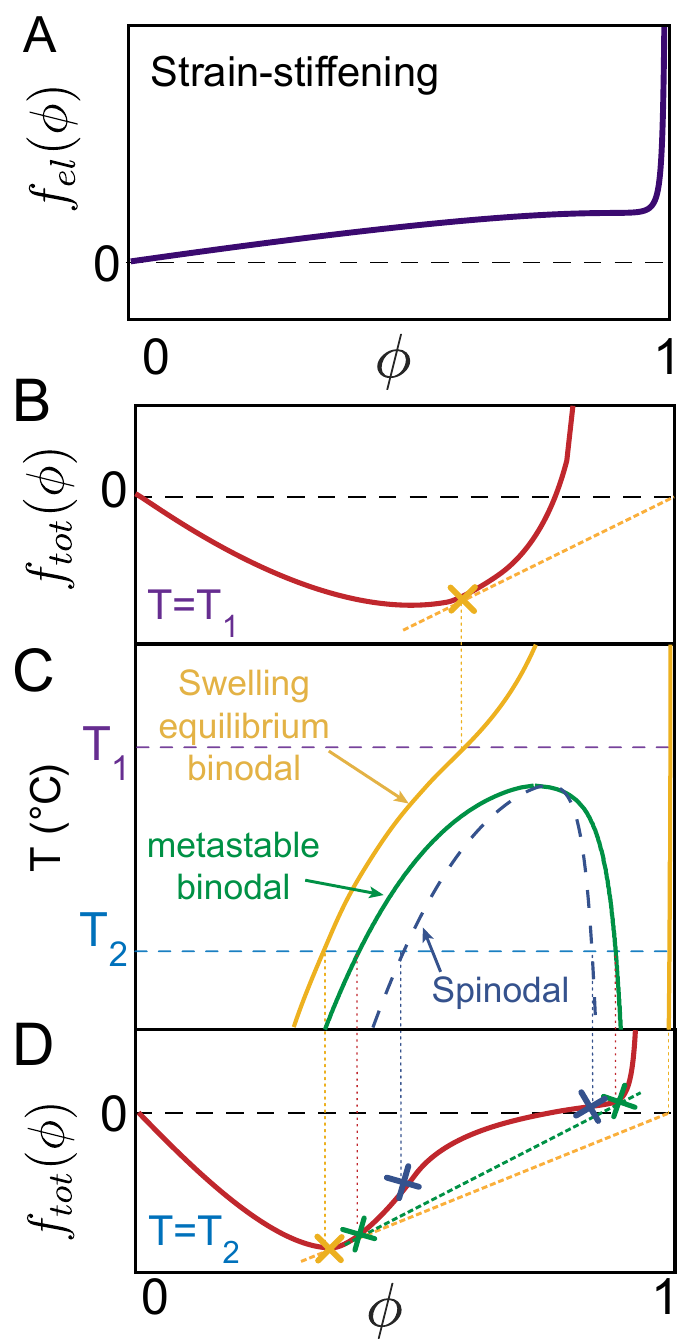}
	\caption{Effect of strain-stiffening on the classical FH free energy of mixing. A) Schematic of the free energy for the Arruda-Boyce model. B, D) Schematic of the total free energy of mixing for strain-stiffening at different temperatures, $T_1$ and $T_2$. C) Schematic of phase diagram that results from A, D.}
	\label{fig:ab_phase_diagram}
\end{figure}

When we introduce strain-stiffening into the elastic network, we obtain a phase diagram (see Figure \ref{fig:ab_phase_diagram}) which much more closely resembles the EMPS phase diagram (Figure 3A,B).
Now, we use the Arruda-Boyce strain-energy density, Eq. (\ref{eq:W_AB}) to obtain the elastic free energy-density, $f_\mathrm{el}^\mathrm{ab}(\phi)$  \cite{okumura2018ultimate}:
\begin{equation}
    f_\mathrm{el}^\mathrm{ab}=\bar{\mu}(1-\phi)\frac{\sum_{i=1}^5 \frac{3^i\alpha_i}{\lambda_m^{2i-2}}((1-\phi)^{-2i/3}-1)}{\sum_{i=1}^5 \frac{i\alpha_i}{\lambda_m^{2i-2}}3^{i-1}},
    \label{eq:f_AB}
\end{equation}
where $\bar{\mu}=\mu /(2k_{\textrm{B}}T/v_s)$, as before.\\

The typical form of $f_\mathrm{el}^\mathrm{ab}$ is shown in Figure \ref{fig:ab_phase_diagram}A. 
While the free energy density decreased monotonically in the neo-Hookean case, there is now a divergence in the elastic energy density as the chains reach their locking stretch: $\sqrt{I_1/3}\rightarrow \lambda_m$. This occurs when $\phi \rightarrow (\lambda_m^3-1)/\lambda_m^3$.

At high temperature (Figure \ref{fig:ab_phase_diagram}B), $f_\mathrm{tot}$ is convex. 
A single-tangent construction (dashed yellow line) reveals the macroscopic swelling equilibrium of a swollen network with pure solvent (yellow cross).
At lower temperatures (Figure \ref{fig:ab_phase_diagram}D), $f_\mathrm{tot}$ becomes partially concave.
As before, a single-tangent construction yields the macroscopic swelling equilibrium (yellow cross).
However, now a double-tangent construction (green dotted line) also reveals the coexistence of two partially-swollen phases (green crosses).
These states have a higher energy than the macroscopic swelling equilibrium (the green line is above the yellow line).
Thus, the partially-swollen phases are metastable with respect to deswelling to the macroscopic swelling equilibrium.
However, when the system is thermally quenched, the microphase separation into partially-swollen states is kinetically favored, since it can be achieved through rapid exchange of solvent between microscopic domains on a timescale $\tau_{\mu PS}\simeq R^2/D$, where $D$ is a diffusion coefficient and $R\simeq\mu\textrm{m}$ is the characteristic domain size. 
To reach the macroscopic swelling equilibrium, solvent must be transported across the whole sample, which takes a time $\tau_{ds}\simeq L^2/D$, where $L\simeq \textrm{mm-cm}$ is the sample size. The ratio of timescales $\tau_{ds}/\tau_{\mu PS}\sim10^6\textrm{-}10^8$ can be enormous, leading to the persistent EMPS phase.

The form of the phase diagram for a swollen strain-stiffening network is shown schematically in Figure \ref{fig:ab_phase_diagram}C.
It displays  the essential features of the data  in Figure \ref{fig:EMPS_phase_diagrams}.
In particular, we obtain the two phase boundaries, representing where macroscopic phase separation (yellow curve) and metastable microscopic phase separation (green curve) occur.
Thus, strain-stiffening behavior appears to be a key ingredient in capturing the phase behavior of EMPS.

\begin{figure}
	\centering
    \includegraphics[width=\columnwidth]{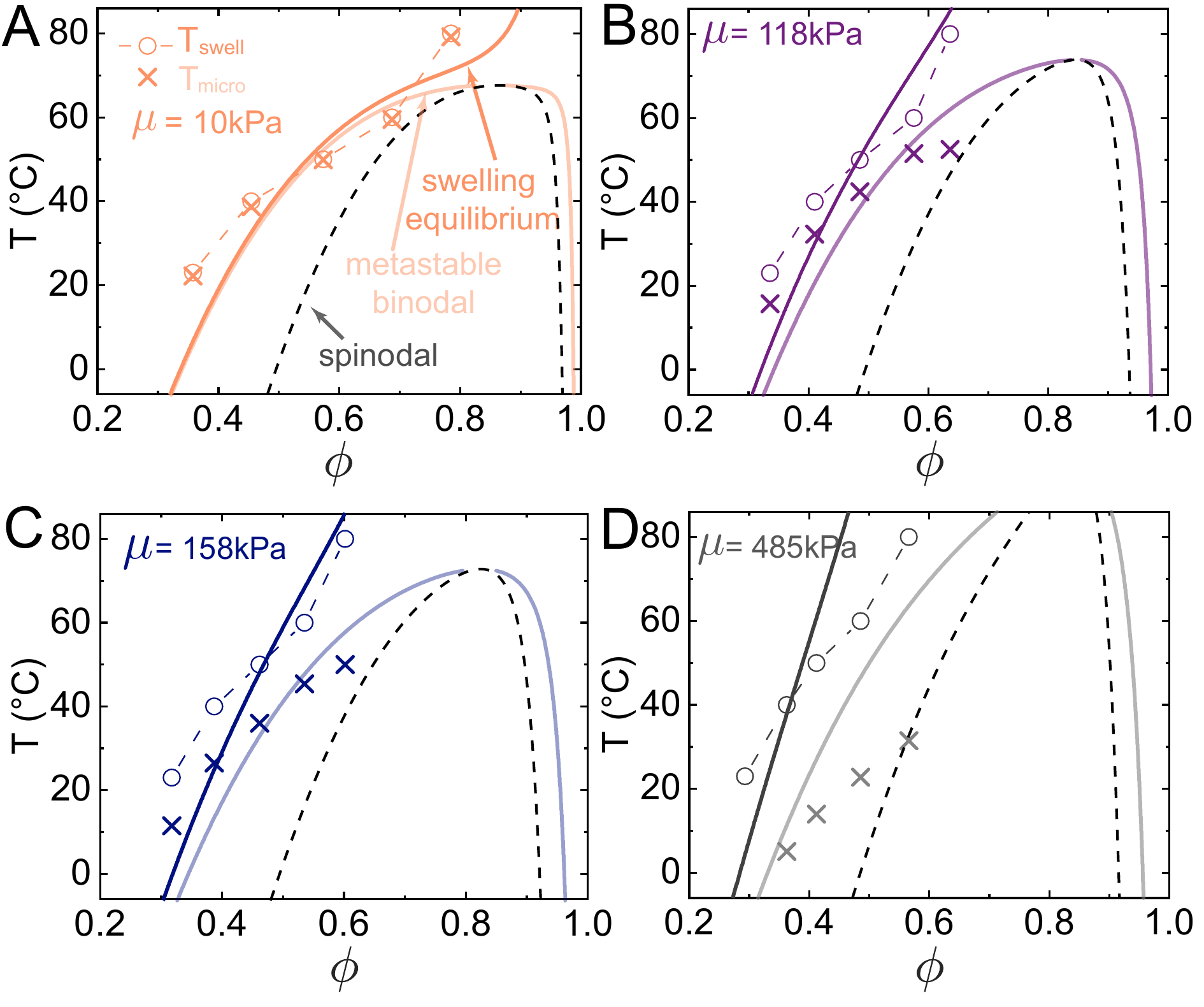}
	\caption{Experimental phase diagrams fitted with the modified Flory-Huggins theory with strain-stiffening. We used values of A) $\mu=10$ kPa and $\lambda_m$=3.7, B) $\mu_m=118$ kPa and $\lambda_m$=3.2, C) $\mu=158$ kPa and $\lambda_m$=3.0 and D) $\mu=$ 485kPa and $\lambda_m$=3.2. Circles and crosses are the experiments and curves are the theoretical swelling equilibrium boundaries (dark color, continuous), metastable binodals (light color, continuous), and  spinodal boundaries (black, dashed).}
	\label{fig:fitted_PS}
\end{figure}

Using this strain-stiffening model of elasticity, we obtain reasonable agreement between calculated phase diagrams and our experimental data, as shown in Figure \ref{fig:fitted_PS}.
There, we have used $\chi(\phi,T)$ from the fit of LLPS of HFBMA and silicone and $\mu$ and $\lambda_B$ from the stress-strain curves of neat ($\phi=0$) silicone elastomers.
Thus, there are no fitting parameters specific to EMPS (see the full expression for $f_{tot}$ in Appendix B).
There is surprisingly good agreement between theory and experiments for both the macroscopic swelling equilibria, and metastable phase separation boundaries (Figure \ref{fig:fitted_PS}).
Deviations between predictions and experiments are largest at high stiffnesses. In that limit, however, the predictions still capture essential trends.

\begin{figure}[ht]
\centering
\includegraphics[width=\columnwidth]{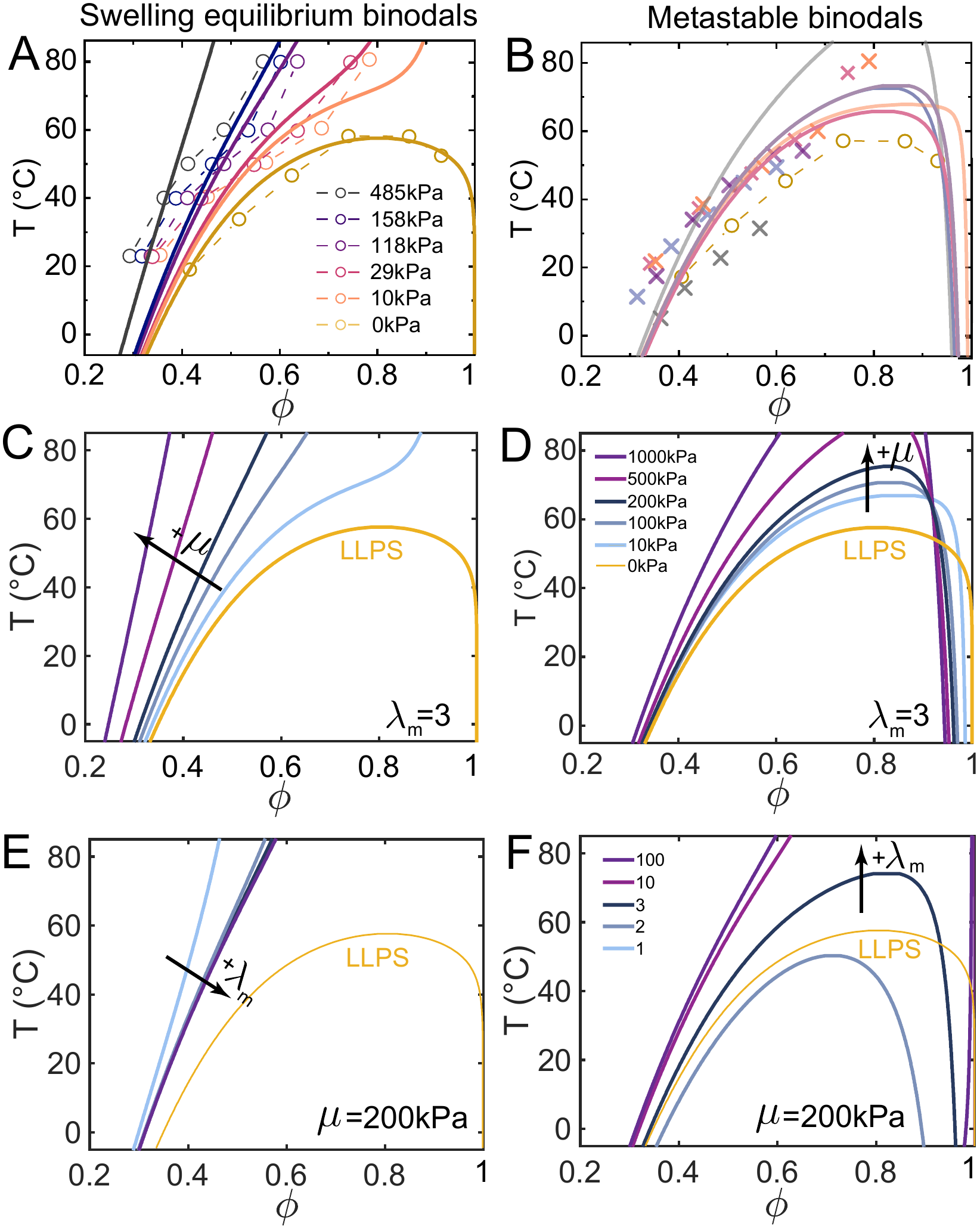}
	\caption{Effect of network mechanics ($\lambda_m$ and $\mu$) on the swelling equilibrium and microphase separation boundaries. A) Experimental swelling equilibrium data (circles) with theory (continuous curves) and B) experimental metastable microphase separation boundaries (crosses) with theory (curves). C-F) Theoretical effect of $\mu$ and $\lambda_m$ on phase diagrams. C,D) $\lambda_m$=3 and $\mu$ varies from 10 to 1000kPa. E,F) Shear modulus is $\mu$=200kPa and $\lambda_m$ varies from 1 to 100. In all figures, yellow circles/curves show the experimental/fitted LLPS phase boundary. Note that for all these calculations, we assume $\alpha$=0.}
	\label{fig:effect_of_elasticity_on_PD}
\end{figure}

Elasticity has different effects on the two phase boundaries in EMPS.
We highlight this by separating the predicted macroscopic swelling equilibrium curves and micro-phase separation curves into two panels (see Figure 8A,B -- yellow curves show LLPS results).
The predicted macroscopic swelling equilibrium is quite strongly affected by increasing network stiffness (Figure \ref{fig:effect_of_elasticity_on_PD}A): the stiffer the network, the less it swells, especially at higher temperatures.
By contrast, the predicted metastable phase boundary is relatively insensitive to changes in network stiffness (Figure \ref{fig:effect_of_elasticity_on_PD}B).
There is a small overall movement of this phase boundary to higher temperatures with increasing stiffness,  inconsistent with the experimental data.

The elastic properties $\mu$ and $\lambda_m$ have different effects on phase equilibria.
Results for  varying $\mu$ at fixed $\lambda_m$ are shown in Figures \ref{fig:effect_of_elasticity_on_PD}C,D, and for varying $\lambda_m$ at fixed $\mu$ are shown in Figures \ref{fig:effect_of_elasticity_on_PD}E,F.
The results show that the polymer network's stiffness has a significantly larger impact than $\lambda_m$ in the position of the swelling equilibrium curve (see Figures \ref{fig:effect_of_elasticity_on_PD}C,E), suggesting that macroscopic swelling equilibrium is relatively insensitive to strain-stiffening.
On the other hand, the metastable binodals seem to be similarly sensitive to both stiffness and strain-stiffening with a shift towards higher temperatures and slightly lower solvent-rich composition with increasing $\mu$ and $\lambda_m$. 

\subsection{Effect of the osmotic contribution to the elastic energy} 

In deriving the phase diagrams above we made an important assumption about the dependence of the elastic energy on volume changes.
In particular, we chose $\alpha=0$ in Eq.~(\ref{eq:alpha}), which follows the phantom-network theory \cite{james1953statistical}.
We revisit this assumption to study how the value of $\alpha$ influences the results.

The value of  $\alpha$ is still debated in the literature \cite{flory1976statistical}. The three most commonly-found predictions allow us to state upper and lower bounds (e.g. \cite{binder1984phase}).
Firstly, the phantom-network model of James and Guth \cite{james1953statistical} predicts that $\alpha=0$.
Secondly, Flory's affine network theory predicts that $0<\alpha\leq 1$ \cite{flory1976statistical}.
A final approach is to require that the elastic network is stress-free in the as-prepared state, so that $\alpha=1$ (this is typically used for the free energy of compressible, neo-Hookean solids).

\begin{figure}
\centering
\includegraphics[width=\columnwidth]{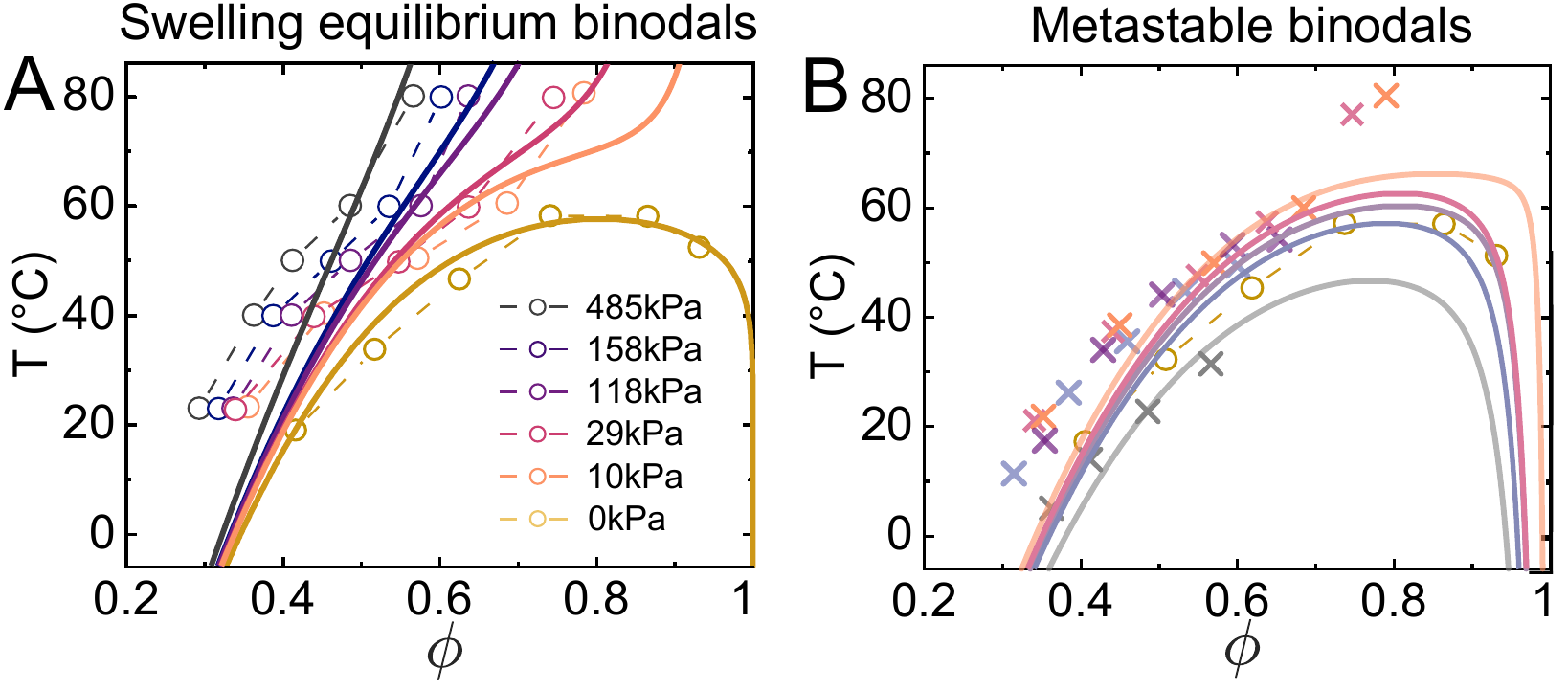}
	\caption{Predicted phase diagrams when an additional volumetric term, $-\mu \log J$, is included in the elastic strain-energy density. A) Swelling equilibrium and B) metastable binodal boundaries. All other parameters are the same as used in Figure \ref{fig:effect_of_elasticity_on_PD}.}
	\label{fig:effect_alpha_on_PD}
\end{figure}

We repeat the phase-boundary calculations with $\alpha=1$ in Figure~\ref{fig:effect_alpha_on_PD}. 
The swelling equilibrium curves with $\alpha=0,1$ are very similar (Figures~\ref{fig:effect_of_elasticity_on_PD}A, \ref{fig:effect_alpha_on_PD}A).
However, the metastable binodals are qualitatively different (Figures~\ref{fig:effect_of_elasticity_on_PD}B, \ref{fig:effect_alpha_on_PD}B).
Increasing $\mu$ now reduces the temperature at which microphase separation occurs (Figure \ref{fig:effect_alpha_on_PD}B).
As a result the prediction of the metastable binodal now agrees much better with the stiffest experimental sample.

\subsection{Structure: Channels vs cavitation}

\begin{figure}[h]
\centering
\includegraphics[width=0.65\columnwidth]{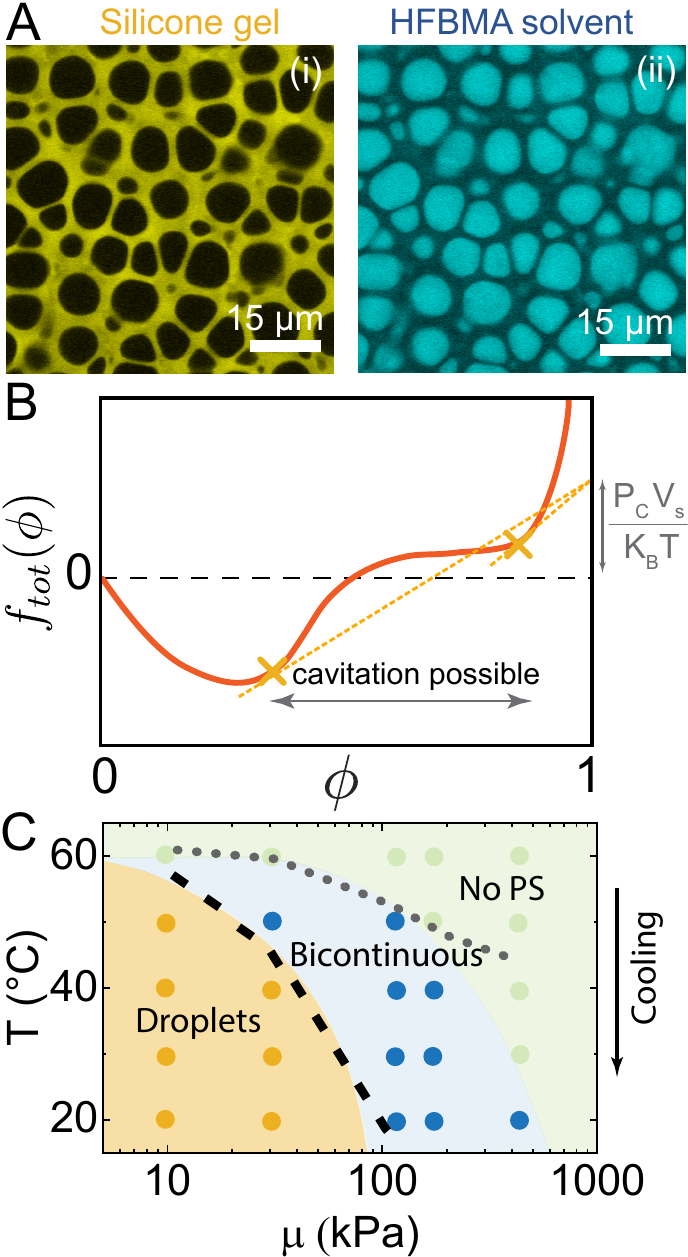}
	\caption{Cavitation in EMPS. A) Raman microscopy images of a cavitated samples where droplets of pure solvent are found surrounded by a matrix. B) Tangent construction for determining the cavitation threshold. C) Experimental data (dots) compared with theory: dashed lines correspond to cavitation threshold (black dashed line) and metastable phase separation boundary (grey dotted line). In this phase diagram samples have been incubated at 60 $^{\circ}$C. 
    }
	\label{fig:cavitation}
\end{figure}

In EMPS experiments, we generally observe the formation of bicontinuous structures at the onset of micro-phase separation (see Figure~\ref{fig:Figure1}B).
The domains are swollen polymer networks, with different concentrations of solvent.
This bicontinuous structure is stable in stiff networks ($\mu >$ 200kPa).
In soft networks, the bicontinous morphology is unstable, and we observe the nucleation and grown of droplets of pure solvent, embedded in swollen networks \cite{fernandez2024elastic}. The composition of the droplets can be quantified with  Stimulated Raman spectroscopy (SRS), see Figure~\ref{fig:cavitation}A, which shows the presence of holes in the silicone network (see right Figure~\ref{fig:cavitation}A), filled in with pure HFBMA solvent (see left image Figure~\ref{fig:cavitation}A).
The latter state can only occur when the phase separation is able to cavitate the silicone networks -- \textit{i.e.} to push open macroscopic, liquid-filled holes in the polymer networks \cite{kim2020extreme,style2018liquid,ronceray2022liquid,vidal2021cavitation,kothari2020effect}. Our theoretical model can be further extended to capture the onset of this cavitation behaviour. 

For cavitation to occur, the excess pressure of solvent that fills any microscopic flaws in the material has to reach a critical value $P_c\sim \mu$ (\textit{e.g.} $P_c=5\mu/2$ for incompressible neo-Hookean elastomers \cite{gent1959internal}).
At this point, the liquid can push open the flaws without an further increase in pressure \cite{style2018liquid}.
Thus, the criterion for observation of droplets is when the solvent in the EMPS system is in equilibrium with pure solvent in the cavities with an excess pressure, $P_c$.
As shown in Appendix A, this condition can be found graphically with an additional single-tangent construction (Figure \ref{fig:cavitation}B), where a tangent line to the $f_\mathrm{tot}$ curve (dashed line) has one end that goes through the point $[\phi=1,f=P_c v_s/(k_{\textrm{B}}T)]$.
Now, the tangent point represents the onset of cavitation, while the end point represents pure solvent at an excess pressure, $P_c$.

With this, we can predict quite accurately when cavitation occurs.
Figure \ref{fig:cavitation}C shows observed microphase separated morphologies in our experimental system.
We saturate polymer networks in the HFBMA solvent at 60$^{\circ}$C and cool them down, and observe when phase separation/cavitation occurs.
For small amounts of cooling is no phase separation (see green area Figure \ref{fig:cavitation}C). For intermediate cooling, we observe microphase separation with a bicontinuous morphology (see blue area Figure \ref{fig:cavitation}C). For deep cooling, or the softest materials, the bicontinuous morphology abruptly changes to droplets (see orange area Figure \ref{fig:cavitation}C). 
We next apply the two tangent constructions (see Figure~\ref{fig:cavitation}B and Figure 6D) to predict the phase separation boundary (dotted curve in Figure \ref{fig:cavitation}C) and cavitation threshold (dashed curve in Figure \ref{fig:cavitation}C).
For the cavitation boundary, we assume that the cavitation threshold in our swollen strain-stiffening material is the same as for an incompressible neo-Hookean network ($P_c=5\mu/2$). Even with this drastically simplifying assumption we find a remarkably good quantitative agreement between the observed experimental regimes and the predicted ones.

\section{Discussion and Conclusions}

We have demonstrated good agreement between experimental phase diagrams and a simple extension of Flory-Huggins theory, incorporating non-linear elasticity.
We have shown that strain-stiffening is an essential ingredient for metastable phase separation in elastic matrices.
By integrating this with models for elastic cavitation, we can rationalize the selection of bicontinuous and droplet morphologies.

While this model successfully captures the phase diagram,  some observations of EMPS remain unexplained.
First,  microphase separation at the metastable binodal forms domains with a well-defined length scale that does not coarsen.
Our model has no mechanism to select this length scale.
Secondly, microphase separation at the metastable phase boundary is continuous -- 
we observe a vanishingly small  contrast between the domains just below the microphase separation phase boundary.
However, the current model suggests that metastable phase separation should display a sharp contrast between the domains, having compositions on the two sides of the metastable binodal curve.

These considerations imply that the model is still missing some important physics.
For example, while we have assumed isotropic elastic stresses in the network, stresses should be very anisotropic near phase interfaces, and this has been shown to be able to arrest coarsening \cite{kothari2020effect}. 
Further, recent theory  has shown that accounting for the mesh-scale heterogeneity of the polymer network \textit{via} non-local elasticity can qualitatively capture the missing features described above \cite{qiang2024nonlocal,oudich2025phase}.
In this last case, further work is needed to extend these predictions to three-dimensions and to build a experimental foundation for non-local elasticity in real polymer networks.

Despite its imperfections, our model suggests practical design principles. 
First, one needs to select a solvent that swells the elastomer enough to feel the non-linearity of the network.
Second, in order to acheive bicontinous morphologies, the elastomer needs to be stiff enought to avoid cavitation. 

We envision several directions for future work. Most importantly, the physical mechanism underlying length-scale selection remains unresolved.  Further, EMPS experiments have been limited to silicone elastomers.  It  will be fruitful to test these concepts in other polymer networks, like hydrogels. Also, elastic phase separation may be able to pin-down the   logartihmic contribution to swelling,  \textit{i.e} the $\alpha$ term (Eq.~\ref{eq:alpha}) \cite{flory1976statistical}. Finally, from a mechanical perspective, we have investigated only one type of strain-stiffening. Strongly-strain stiffening networks, like biopolymers, could show very different behavior.

\section*{Author contributions}
\textbf{Conceptualisation}: CF-R, RWS, ERD, \textbf{Formal Analysis}: CF-R, RWS, SH, SW, PDO, ERD \textbf{Investigation}: CF-R,
\textbf{Methodology}: CF-R, SH, RWS, \textbf{Resources}: ERD, \textbf{Software}: RWS, SH, \textbf{Supervision}: RWS, ERD, \textbf{Validation}: SW, PDO, \textbf{Visualisation}: CF-R, RWS, \textbf{Writing - Original Draft}: CF-R, RWS, \textbf{Writing - Review \& Editing}: CF-R, RWS, PDO, ERD.

\section*{Conflicts of interest}
There are no conflicts to declare.

\section*{Data availability}
Data for this article, including data for the mechanical characterization and code to generate phase diagrams are available at the Figshare repository at https://doi.org/10.6084/m9.figshare.29153642.

\section*{Acknowledgements}
We thank Sanat Kumar for useful discussions. We thanks Justine Kush and Dorothea Pinotsi from ScopeM for help during the SRS experiments. CF-R acknowledges funding from the ETH Zürich Postdoctoral Fellowship. PDO and SW thank the Ives Foundation and Georgetown University for support. 

\section*{Materials \& Methods}

\subsection*{Fabrication of polymer matrices}
We prepare pure PDMS matrices by mixing PDMS chains (DMS-V31, Gelest, $M_w$=28000g/mol) with a crosslinker (HMS-301, Gelest $M_w$=1900g/mol) and a platinum-based catalyst (SIP6831.2, Gelest) (see full recipe in \cite{Style2015}). We typically prepare two parts: Part A which consists of PMDS chains with 0.05~wt\% of catalyst and Part B which consists of PDMS chains with 10~wt\% of crosslinker. The stiffness of the resulting matrix depends on the mass ratio between part A and part B (typically ranging from 3:1 to 9:1), while keeping the catalyst concentration constant (0.0019\% in volume). Once the different parts are thoroughly mixed together, we pour the mixture into a petri dish, degas it in vacuum, and finally cure it at 60$^{\circ}$C for approximately 6 days. After curing, the resulting PDMS elastomer is carefully removed from the petri dish and cut into rectangular pieces ($\sim$1cm$\times$2cm$\times$0.5cm). 

\subsection*{EMPS and LLPS experiments}

PDMS pieces are transferred into a bath of heptafluorobutyl methacrylate (HFBMA, Apollo scientific, $M_w$=268.1) ($\sim$1mL HFBMA/0.5g of PDMS), in a 25mL glass bottle. This bath is then typically incubated at $T_{swell } = 60^{\circ}$C in a pre-heated oven during 2.5 days. After the elastomer is saturated with the liquid, the sample is then transferred into a preheated heating stage at 60$^\circ$C (INSTEC), which is coupled to an optical microscope. We then use custom image analysis routines to carefully detect the emergence of phase separation as a function of temperature. For these experiments, we use a 60 $\times$ air objective, to avoid temperature variations in the sample.

Note that experiments with uncrosslinked PDMS are performed by preparing mixtures of PDMS chains and HFBMA liquid in glass bottles, and then following the same steps described above for determining the phase separation boundaries.

\subsection*{Swelling equilibrium experiments}

To measure the equilibrium concentration of HFBMA in PDMS as a function of the swelling temperature, T$_{\rm swell}$, we take 1$\times$2$\times$0.5cm$^3$ PDMS samples and record their original weight (m$_{\rm dry}$). We then incubate the samples into bath of HFBMA liquid ($\sim$1mL) in a well-sealed glass bottle. The samples are left incubating for 2.5-3 days at the desired swelling temperature. After the incubation, we remove the swollen pieces of PDMS from the bath, gently remove the excess of liquid on its surface, and measure their increase in mass with a microbalance (m$_{\rm swollen}$). This extra mass corresponds to the liquid that has diffused inside the PDMS. However, we also note that a fraction of uncrosslinked PDMS leaves the PDMS during the incubation period and goes to the bath of HFBMA. To take this mass of uncrosslinked PDMS into account, we let the HFBMA evaporate from the swollen sample for a period of 24 hours. We then weigh the mass of the resulting dry PDMS (m$_{\rm evap}$). This mass is usually 5-20\% lighter than the original dry mass of PDMS depending on the stiffness of the sample and swelling temperature. The mass fraction of liquid inside the matrix is calculated as $m_{\rm liquid} = m_{\rm swollen}-m_{\rm evap} / m_{\rm swollen} $. The resulting the volume fraction of the liquid, $\phi$, is then calculated as: 

\begin{equation}
\phi  = \frac{m_{\rm liquid} / \rho_{\rm liquid}}{m_{\rm liquid} / \rho_{\rm liquid} + m_{\rm matrix} / \rho_{\rm matrix}}
\end{equation}

Where $\rho_{\rm liquid} = 1.345 \textrm{g/mL}$, $\rho_{\rm matrix} = 0.97 \textrm{g/mL}$ and $m_{\rm matrix} = 1 - m_{\rm liquid}$.
Note that the measured value of $\phi$ corresponds to the binodal curve of system, as $\phi$ is the concentration of one of the coexisting phases at equilibrium fixed at T$_{\rm swell}$. We ensure that the system has reached equilibrium by incubating the sample for $\sim$ twice as long as the time when we no longer record a mass increase of the swollen elastomer, $m_{swollen}$ (after $\sim$ 24 hours).

\subsection*{Uniaxial tensile tests and fitting parameters}

To measure the non-linear behaviour of the pure, unswollen silicone networks, we prepare cylindrical silicone samples by using Teflon tubes with an internal diameter of 1.5mm as molds. After pouring the mixture of PDMS chains and crosslinker into the tubes, we transfer them into a 60$^\circ$C oven for a week. After fully curing the materials, we carefully peal off the teflon tubing, and cut the resulting PDMS cylinders into lengths of 5cm long pieces. We then glue the end of the tubes into rectangular pieces of PDMS of the same stiffness, using a drop of the PDMS mixture as a glue. The rectangular ends enable an easier clamping for the mechanical test. We cure the whole construct over two days at 60 deg$^\circ$C. 

 We next perform uniaxial tests on our samples using a tensile testing machine (Stable Micro Systems), where we record the engineering strain-stress curves until failure. All mechanical tests were performed in air, at room temperature, and at elongation speeds of 0.05mm per second.

 Using the Arruda-Boyce model \cite{arruda1993three}, we fit the strain-stress curves for multiple samples with five different crosslinking densities and obtain the  fitted values shown in Table~\ref{table}.
 Each value reports average results for three replicates -- except the samples with 7.4 mol\% crosslinker, where two specimens were measured. All values of $\mu$ deviated from the mean by less than $\sim 15\%$, and values of $\lambda_{m}$ by less than $\sim 20\%$.

 \begin{table}[h]
\small
  \caption{\ Arruda-Boyce fitted values of the shear modulus $\mu_{AB}$ and maximum microscopic polymer elongation $\lambda_m$, for pure, unswollen silicone networks as a function of crosslinker density \label{table}}
  \begin{tabular*}{0.48\textwidth}{@{\extracolsep{\fill}}ccc}
    \hline
    Crosslinker density (mol\%) & $\mu_{AB}$ (kPa) & $\lambda_m$ (-) \\
    \hline
    12.9 & 485.3
    & 3.2 \\
    10.3 & 158.4 & 3.0 \\
    7.4 & 117.7 & 3.2 \\
    6.4 & 29.8 & 2.6 \\
    5.2 & 9.8 & 3.7 \\
    \hline
  \end{tabular*}
  \label{table}
\end{table}

\subsection*{Calculating phase diagrams}
We use the convex-hull technique described by \cite{mao2019phase} to calculate phase diagrams.
In brief, this finds concave sections of a free-energy curve by comparing the curve with its convex hull.
Example code is given as a Supplement to this paper.

\section*{Appendix A: Conditions for phase equilibrium}

Following Doi \cite{doi2013soft}, we can write the Gibbs free energy of a two component, solvent/polymer system as
\begin{equation}
    G=PV+\frac{k_{\textrm{B}}T}{v_s}f(\phi,T)V,
\end{equation}
where $P$ is the (constant) bulk pressure, $V$ is the (constant) volume, and $f$ is the non-dimensional Helmholtz free energy per unit volume -- the same as $f_\mathrm{tot}$ from the manuscript.

The chemical potential of the solvent in this system is 
\begin{equation}
    \mu^*=\left(\frac{\partial G}{\partial N_s}\right)_{V_p,T}=v_sP+k_{\textrm{B}}T\left(f+(1-\phi)\frac{\partial f}{\partial \phi}\right),
\end{equation}
where we have used that $V=v_sN_s+V_p$, and $\phi=v_sN_s/V$, where $N_s$ is the number of solvent molecules in the system, and $V_p$ is the volume of polymer.
We can also define an \textit{exchange chemical potential}, as the energy change caused by adding a solvent molecule, and removing the same volume of polymer:
\begin{equation}
    \mu^*_\mathrm{ex}=\left(\frac{\partial G}{\partial N_s}\right)_{V,T}=k_{\textrm{B}}T\frac{\partial f}{\partial \phi}.
\end{equation}

For equilibrium between two phases that both contain a mixture of polymer and solvent, both $\mu^*$ and $\mu^*_\mathrm{ex}$ must be the same in the two phases.
The latter condition implies that the two equilibrium phases $(\phi_1,\phi_2)$ must be at points with the same tangent slope.
The two conditions can also be combined to show that
\begin{equation}
    \frac{f(\phi_2)-f(\phi_1)}{\phi_2-\phi_1}=\left.\frac{\partial f}{\partial\phi}\right|_{\phi_1}=\left.\frac{\partial f}{\partial\phi}\right|_{\phi_2}.
\end{equation}
Geometrically, both of these conditions are then satisfied by the double tangent construction.

When a polymer network is in equilibrium with pure solvent, we can no longer exchange polymer between the two phases. So we lose the requirement of having the same $\mu^*_\mathrm{ex}$ in the two phases.
Instead we only require that the chemical potential of pure solvent, at a pressure $P$: $\mu^*_s=Pv_s$ is equal to the chemical potential of the solvent in the swollen polymer network.
This gives that
\begin{equation}
    f+(1-\phi)\frac{\partial f}{\partial \phi}=0.
\end{equation}
This is satisfied by a single-tangent construction  with a tangent line that passes through the point $(\phi=1,f=0)$.

Finally, for the cavitation condition, we  need equilibrium between a swollen polymer network and a pure solvent with excess pressure $P_c$ (\textit{i.e.} with total pressure $P+P_c$).
In this case, we again only require equality of $\mu^*$ between the two phases.
The chemical potential of the solvent is $\mu^*_s=Pv_s+P_cv_s$.
Thus, cavitation occurs when
\begin{equation}
    f+(1-\phi)\frac{\partial f}{\partial \phi}=\frac{v_sP_c}{k_{\textrm{B}}T}.
\end{equation}
This is satisfied by a single-tangent construction  with a tangent line that passes through the point $(\phi=1,f=v_sP_c/k_{\textrm{B}}T)$.
\newline
\section*{Appendix B: Full expressions for EMPS free energies}

The total dimensionless free-energy density in a swollen neo-Hookean gel is
\begin{align}
    f_\mathrm{tot}(\phi,T)=& \phi \log\phi + \left[A+\frac{B}{T} + C\phi\right] \phi (1-\phi)\\ \nonumber
    &+3\bar{\mu}\left[\left(1-\phi\right)^{1/3}-(1-\phi)\right] \\ \nonumber
    &+\frac{\alpha v_s}{k_bT}(1-\phi)\log(1-\phi).
\end{align}
The total dimensionless free-energy  density  in a swollen Arruda-Boyce gel is
\begin{align}
    f_\mathrm{tot}(\phi,T)=& \phi \log\phi + \left[A+\frac{B}{T} + C\phi\right] \phi (1-\phi)\\ \nonumber
    &+\bar{\mu}(1-\phi)\frac{\sum_{i=1}^5 \frac{3^i\alpha_i}{\lambda_m^{2i-2}}((1-\phi)^{-2i/3}-1)}{\sum_{i=1}^5 \frac{i\alpha_i}{\lambda_m^{2i-2}}3^{i-1}}\\ \nonumber
    &+\frac{\alpha v_s}{k_bT}(1-\phi)\log(1-\phi).
\end{align}
All constants are given in the main text.





%

\end{document}